\documentstyle[twocolumn,aps,epsf]{revtex}

\begin{document}

\newcommand{\bi}[1]{\mbox{\boldmath$#1$}}

\draft

\title{Visualization of correlation cascade in spatio-temporal chaos
  using wavelets}

\author{Hiroya Nakao${^1}{^*}$, Tsuyoshi Mishiro${^2}$, and Michio
  Yamada${^3}$}

\address{${^1}$RIKEN Brain Science Institute, 2-1 Hirosawa, Wako,
  Saitama 351-0198, Japan,\\ 
  ${^2}$Department of Physics, Graduate School of Sciences, Kyoto
  University, Kyoto 606-8502, Japan,\\ 
  ${^3}$Graduate School of Mathematical Sciences, University of Tokyo,
  3-8-1 Komaba, Meguro,\\ Tokyo 153-8914, Japan.}

\date{June 7, 2000}

\maketitle

\begin{abstract}
  We propose a simple method to visualize spatio-temporal correlation
  between scales using wavelets, and apply it to two typical
  spatio-temporally chaotic systems, namely to coupled complex
  Ginzburg-Landau oscillators with diffusive interaction, and those
  with non-local interaction.
  Reflecting the difference between underlying dynamical processes,
  our method provides distinctive results for those two systems.
  Especially, for the non-locally interacting case where the system
  exhibits fractal amplitude patterns and power-law spectrum, it
  clearly visualizes the dynamical cascade process of spatio-temporal
  correlation between scales.
  \\
  \\
  $^{*}$ Corresponding author.
  \\
  \ \ E-mail: nakao@mns.brain.riken.go.jp
  \\
\end{abstract}


\section{Introduction}

It has long been a subject of discussion how to capture and describe
complex dynamical processes of strongly correlated many-body systems
that are ubiquitous in nature.
Probably one of the most important problems of such kind at present is
the one faced in neuroscience. In what way biological information
processing is encoded into spikes of neurons in the brain is the
greatest mystery of present neuroscience. In order to capture subtle
correlation between activities of neurons, information-theoretical
techniques have been developed and applied to experimental signals
[Rieke {\it et al.}, 1998].
%%\cite{Rieke}.

On the other hand, we may mention Richardson's cascade picture of
fully-developed fluid turbulence as a successful example of capturing
the essence of a strongly correlated system, which stated that large
eddies gradually fragment into smaller eddies, and eventually
dissipate due to viscosity. Apart from the difficult problem of
intermittency, this picture was mathematically formulated into the
Kolmogorov 1941 theory, which gave the famous $-5/3$ energy spectrum
[Frisch, 1995].
%%\cite{Frisch}.
%%
This success was due to the fact that in fluid turbulence,
fortunately, a simple hierarchy in spatial scale (or equivalently, in
temporal / energy scales) is formed as a result of complex nonlinear
dynamics of a number of modes.
Since the success in fluid turbulence, attempts have been made to
capture the dynamics of other kinds of many-body systems such as
spatio-temporal chaos [Ikeda \& Matsumoto, 1989],
%%\cite{Ikeda},
chaotic systems with large degrees of freedom [Kaneko, 1994],
%%\cite{Kaneko},
and stock markets [Arn\'eodo {\it et al.}, 1998a],
%%\cite{Arneodo1},
based on the picture that energy, or more generally information or
causality, flows from scales to scales.

In analyzing space-scale properties of various kinds of signals, the
wavelet transform is expected to be a useful tool due to its localized
nature both in space and scale, and many applications have been made
[Mallat, 1999].
%%\cite{Mallat}.
%%
Of course, there are many studies that tried to capture the cascade
process of fully-developed turbulence directly using wavelets [Argoul
{\it et al.}, 1989; Yamada \& Ohkitani, 1991; Arn\'eodo {\it et al.},
1998b].
%%\cite{Argoul,Yamada,Arneodo2}.
But most of the studies treated one-dimensional time sequences of the
velocity field measured at a single point in fluid, and rather a
limited number of studies actually investigated the temporal evolution
of correlation between scales of spatially extended fields [Toh,
1995].
%%\cite{Toh}.

In this paper, we propose a simple visualization method of such
spatio-temporal correlation between scales of spatially extended
fields using wavelets, and apply it to two typical spatio-temporally
chaotic systems. As such chaotic systems, we investigate ordinary
diffusively coupled complex Ginzburg-Landau oscillators [Kuramoto,
1984; Bohr {\it et al.}, 1998]
%%\cite{Kuramoto1,Bohr}
and somewhat unusual non-locally coupled complex Ginzburg-Landau
oscillators [Kuramoto, 1995].
%%\cite{Kuramoto2}.
While the former system exhibits spatially smooth amplitude patterns,
the latter system exhibits fractal spatial patterns. Therefore, the
dynamical processes behind those systems are expected to differ from
each other significantly, and it is interesting to see whether their
difference can be detected by our method or not.

\section{Coupled complex Ginzburg-Landau oscillators}

The ordinary diffusively (i.e., locally) coupled complex
Ginzburg-Landau equation [Kuramoto, 1984; Bohr {\it et al.}, 1998]
%%\cite{Kuramoto1,Bohr}
is given by
\begin{equation}
  \dot{W}(x, t) = W - (1 + i c_2) |W|^2 W + D (1 + i c_1) \nabla^2 W,
\end{equation}
which can be derived, for example, from equations of oscillatory media
in the vicinity of their Hopf bifurcation points by the
center-manifold reduction technique.
Here, $W(x,t)$ is a complex amplitude of an oscillator at position $x$
and at time $t$, $c_1$ and $c_2$ are real parameters, and $D$ is a
diffusion constant.
It is well known that this equation exhibits spatio-temporal chaos in
some appropriate parameter region.
We call this equation ``LCGL equation'' hereafter.

In this paper, we treat only spatially one-dimensional cases. We fix
the length of the system at $L=1$, and assume a periodic boundary
condition. The diffusion constant is set at $D=0.0035$, and the
parameters are fixed at $c_1=-2$ and $c_2=2$. With these values, the
spatially uniform solution of the LCGL equation is unstable, and the
system exhibits spatio-temporal chaos.
Typical time scale of the system is roughly estimated as the time
needed for an uncoupled free oscillator to go around its limit cycle,
and is given by $2 \pi / c_2 \simeq 3$.
The numerical simulation was done in wavenumber space using
$N=2^{10}-2^{14}$ modes by the pseudo-spectral method with a time step
of $0.01$ (Euler integration).

\begin{figure}[htbp]
  \begin{center}
    \leavevmode
    \epsfxsize=0.4\textwidth
    \epsfbox{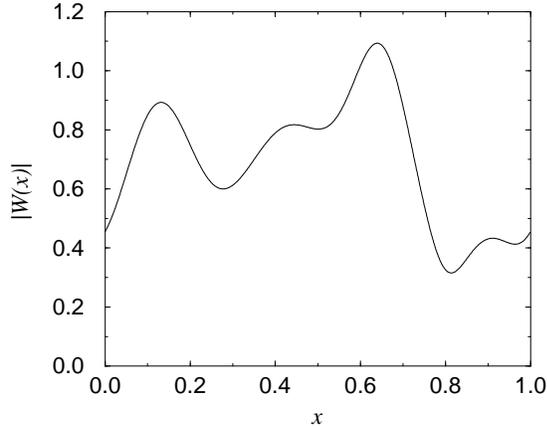}
    \caption{Snapshot of the amplitude field $|W(x,t)|$ of the LCGL equation.
      $N=2^{10}$ modes are used in the numerical calculation.}
    \label{Fig:01}
  \end{center}
\end{figure}

\begin{figure}[htbp]
  \begin{center}
    \leavevmode
    \epsfxsize=0.4\textwidth
    \epsfbox{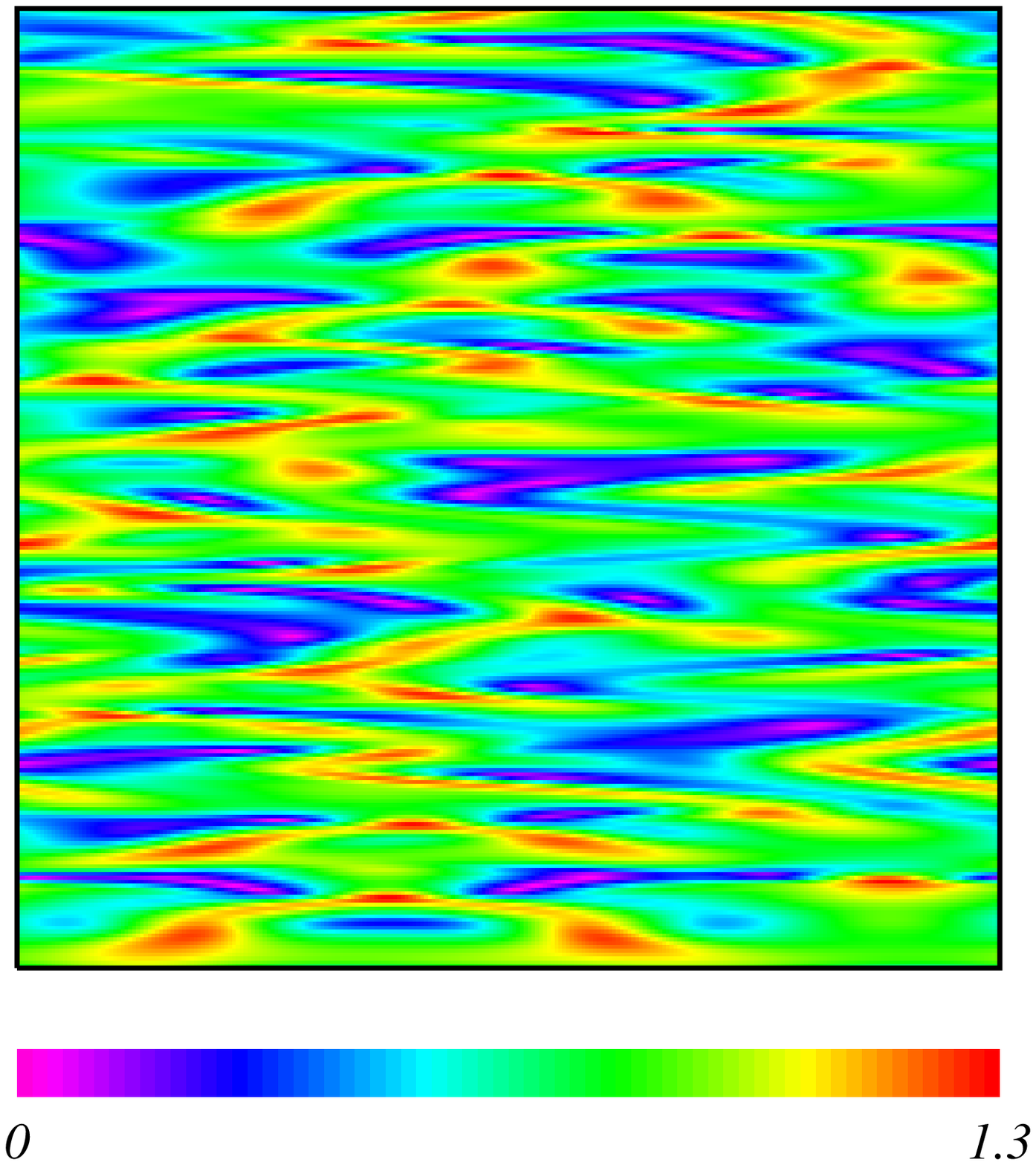}
    \caption{Evolution of the amplitude field of the LCGL equation.
      The horizontal axis indicates the spatial coordinate $x \; (0
      \leq x < 1)$, and the vertical axis the time $t \; (0 \leq t <
      125)$. Time increases upwards. Initial transient is discarded.}
    \label{Fig:02}
  \end{center}
\end{figure}

\begin{figure}[htbp]
  \begin{center}
    \leavevmode
    \epsfxsize=0.4\textwidth
    \epsfbox{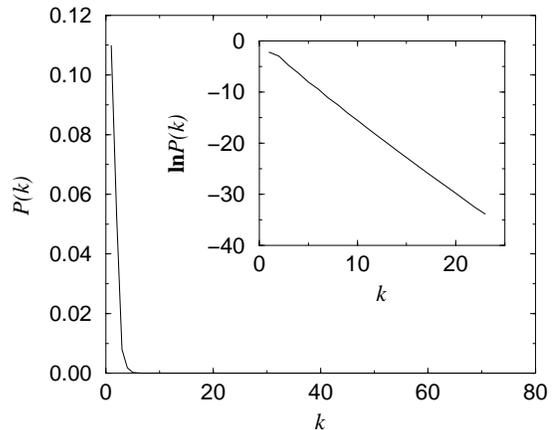}
    \caption{Power spectrum of the amplitude field of the LCGL equation.
      The inset shows semi-log plot of the same spectrum.}
    \label{Fig:03}
  \end{center}
\end{figure}

Due to the existence of a diffusion term, the solution of the LCGL
equation necessarily possesses a characteristic minimal length scale,
below which fluctuations are strongly depressed.
Thus, the amplitude pattern $|W(x,t)|$ of the solution has a smoothly
modulated shape as displayed in Fig.~\ref{Fig:01}.
The shortest wavelength determined by the diffusion constant is about
$\sqrt{D} \simeq 0.06 \simeq 1/16$. For the sake of comparison, this
value is equated with the coupling length of the non-locally coupled
system that we explain later.
Figure~\ref{Fig:02} displays temporal evolution of the amplitude
pattern, and Fig.~\ref{Fig:03} shows its power spectrum. Since
short-wavelength components are strongly depressed, the power spectrum
decays exponentially.


%%\vspace{1cm}
\newpage

As another spatio-temporally chaotic system, we investigate a
different type of coupled complex Ginzburg-Landau oscillators, which
was first introduced by Kuramoto [1995].
%%\cite{Kuramoto2}.
Instead of a diffusive interaction, it has a non-local interaction;
each oscillator feels a non-local mean field of other oscillators
through a kernel $g(|x|)$, which is a decreasing function of
$|x|$. Hereafter we use $g(|x|) = g_0 \exp(-|x| / \gamma)$ as the
kernel, where $\gamma$ gives the coupling range. $g_0$ is some
appropriate normalization constant, and assumed to be $2 \gamma$.
The equation for this system of non-locally coupled oscillators is
given by
%%
%\begin{equation}
%  \dot{W}(x, t) = W - (1 + i c_2) |W|^2 W
%  + K (1 + i c_1) \int dx' g(|x'-x|) \left[ W(x') - W(x) \right],
%\end{equation}
%%
\begin{eqnarray}
  & \dot{W}&(x, t) = W - (1 + i c_2) |W|^2 W \cr \cr
  & & + K (1 + i c_1) \int dx' g(|x'-x|) \left[ W(x', t) - W(x, t) \right],
\end{eqnarray}
which we call ``NCGL equation'' hereafter. This equation can be
derived, for example, from a model of oscillatory biological cells
that are interacting through some diffusive chemical substance with a
finite decay rate.

Although the difference between this NCGL equation and the LCGL
equation is only the last interaction term, it was shown that this
NCGL equation exhibits spatio-temporal chaos with remarkably different
features from the LCGL equation, such as fractal amplitude patterns,
power-law spatial correlation, and power-law spectrum.
We again fix the length of the system at $L=1$, and assume a periodic
boundary condition. The coupling range is fixed at $\gamma=1/16$,
which is equal to the shortest characteristic length $\sqrt{D} \simeq
1/16$ of the LCGL equation. (More precisely, by assuming the
smoothness of the amplitude field $W(x',t)$ in the interaction term
and expanding it up to the second order in $x'-x$ around $x$, we
obtain an effective diffusion constant $D_{eff} = K \gamma^2$ for the
NCGL equation.  However, the smoothness of the amplitude field is not
always guaranteed, hence this correspondence is only formal.)
The parameter values are again set at $c_1 = -2$ and $c_2 = 2$. With
these values, the uniform solution of the system is always unstable,
and the system behaves in a chaotic manner.
Further, the behavior of the system strongly depends on the remaining
coupling strength $K$; the amplitude pattern of the system is smooth
for large $K$ ($\sim 1.5$), while almost discontinuous for small $K$
($\sim 0.5$). Between these values, there exists a parameter region
where the amplitude pattern is fractal, and its spatial correlation
exhibits power-law behavior.
We fix $K=0.9$ hereafter, where the system is exactly in this
``anomalous'' spatio-temporally chaotic regime.
The numerical simulation was done in real space using
$N=2^{10}-2^{14}$ oscillators by the 4th-order Runge-Kutta method with
a time step of $0.05$.

\begin{figure}[htbp]
  \begin{center}
    \leavevmode
    \epsfxsize=0.4\textwidth
    \epsfbox{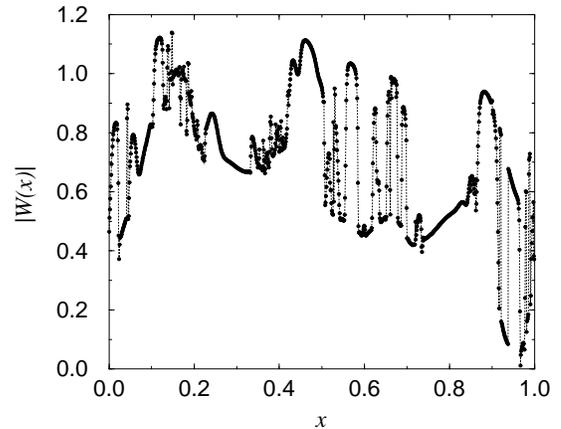}
    \caption{Snapshot of the amplitude field $|W(x,t)|$ of the NCGL
      equation. $N=2^{10}$ oscillators are used in the numerical
      calculation.}
    \label{Fig:04}
  \end{center}
\end{figure}

\begin{figure}[htbp]
  \begin{center}
    \leavevmode
    \epsfxsize=0.4\textwidth
    \epsfbox{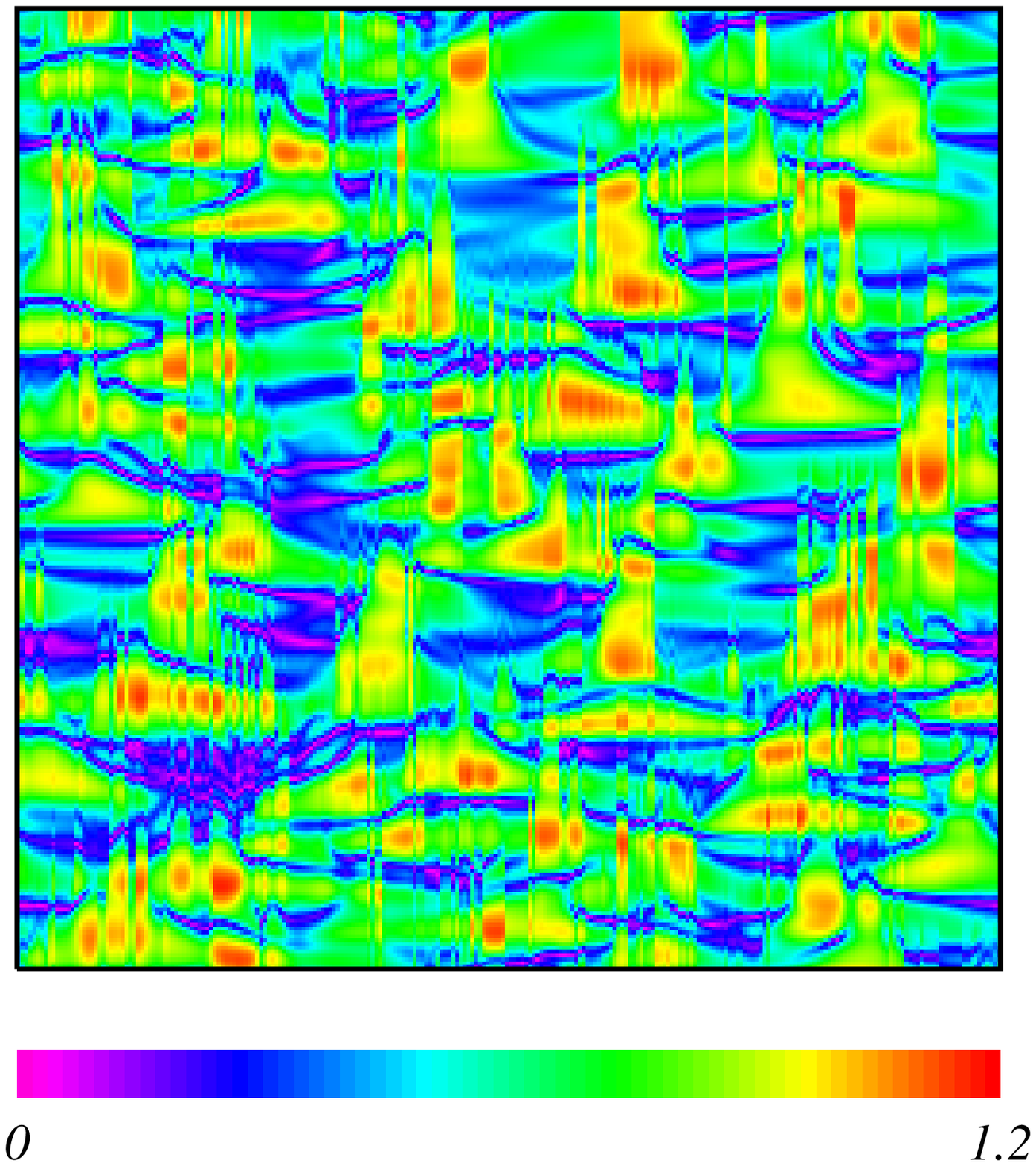}
    \caption{Evolution of the amplitude field of the NCGL equation.
      The horizontal axis indicates the spatial coordinate $x \; (0
      \leq x < 1)$, and the vertical axis the time $t \; (0 \leq t <
      125)$. Time increases upwards. Initial transient is discarded.}
    \label{Fig:05}
  \end{center}
\end{figure}

\begin{figure}[htbp]
  \begin{center}
    \leavevmode
    \epsfxsize=0.4\textwidth
    \epsfbox{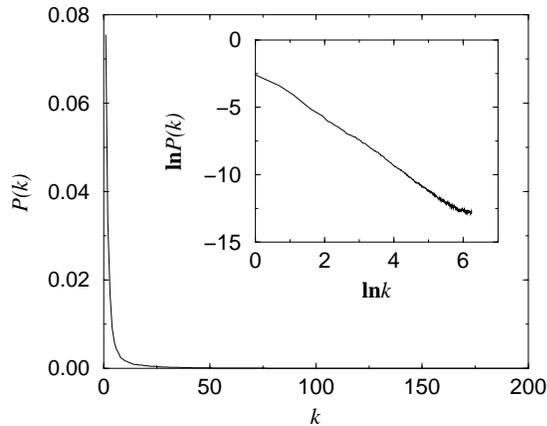}
    \caption{Power spectrum of the amplitude field of the NCGL equation.
      The inset shows log-log plot of the same spectrum.}
    \label{Fig:06}
  \end{center}
\end{figure}

Figure~\ref{Fig:04} displays a typical snapshot of the solution of the
NCGL equation, and Fig.~\ref{Fig:05} shows its temporal evolution.
In contrast to the case of the LCGL equation, the amplitude pattern is
not completely smooth, but is composed of patches of smooth coherent
regions and strongly disordered regions.
The power spectrum displayed in Fig.~\ref{Fig:06} also indicates the
difference clearly. It exhibits power-law decay rather than the
exponential decay in the case of the LCGL equation, which implies the
strong anomaly of the amplitude field.
Actually, it was shown that the amplitude pattern of the NCGL equation
is fractal, and its fractal dimension varies with the coupling
strength $K$.
These properties are essentially the results of the absence of length
scale shorter than the coupling range $\gamma$, and can be explained
using a simple multiplicative stochastic model to a certain extent
[Kuramoto \& Nakao, 1996].
%%\cite{Kuramoto3}.


Thus, the difference in the interaction term greatly changes the
amplitude pattern of the system, even though the oscillators and
characteristic length scales are the same.
Especially, the power-law behavior of the power spectrum reminds us of
the $-5/3$ energy spectrum of fluid turbulence. (However, the exponent
changes with the coupling strength $K$ for our NCGL equation.)
The power-law behavior of the energy spectrum in fluid turbulence is
related to the cascade process of breakdown of vortices. Thus, we may
naively expect that our NCGL equation also possesses some kind of
cascade process. Of course, the NCGL equation is strongly dissipative
and there is no such conservative quantity like energy or enstrophy.
However, we still expect a cascade process of some quantity from
long-wavelength components to shorter-wavelength components, which may
be called causality or information, or merely correlation.
In the following part of this paper, we develop a simple method to
visualize the spatio-temporal correlation between fluctuations at
different scales, in order to prove the existence of such cascade
process.

\section{Wavelet-based correlation analysis}

\begin{figure}[htbp]
  \begin{center}
    \leavevmode
    \epsfxsize=0.4\textwidth
    \epsfbox{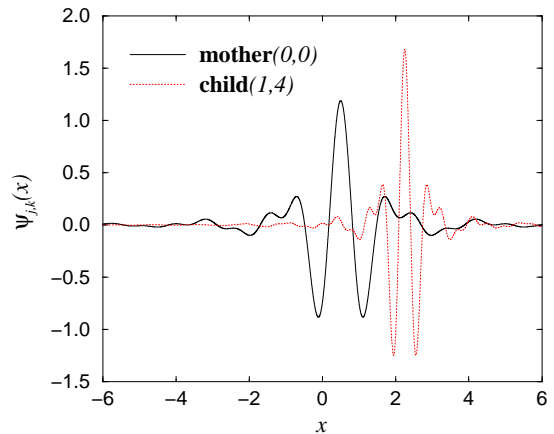}
    \caption{Meyer's wavelets; mother wavelet ($j=0$, $k=0$)
      and one of its child wavelet ($j=1$, $k=4$).}
    \label{Fig:07}
  \end{center}
\end{figure}

Let us consider decomposing the spatial pattern into various scales
using wavelets, and analyzing the temporal correlation between them.
We decompose a spatial pattern $V(x)$ using an orthonormal wavelet
basis as
\begin{equation}
  V(x) = \sum_{j=-\infty}^{\infty} \sum_{k=-\infty}^{\infty} a_{j,k}
  \psi_{j,k}(x),
  \label{Eq:WaveletDecomposition}
\end{equation}
where $a_{j,k}$ is an expansion coefficient, and $\psi_{j,k}(x)$ is a
child wavelet generated from the mother wavelet $\psi(x)$ by
translation and dilation as
\begin{equation}
  \psi_{j,k}(x) = \sqrt{2^j} \psi (2^j x - k),
\end{equation}
where $j$ is a scale parameter, and $k$ is a translation parameter.
The child wavelets are mutually orthogonal as
\begin{equation}
  \langle \psi_{j,k}, \psi_{j',k'} \rangle = \int_{-\infty}^{\infty}
  \psi_{j,k}(x') \psi^{*}_{j',k'}(x') dx' = \delta_{j,j'}
  \delta_{k,k'},
\end{equation}
and form a complete orthonormal set.
Since the child wavelet $\psi_{j,k}(x)$ is localized both in space and
scale roughly at position $k / 2^j$ and scale $1 / 2^j$, the expansion
coefficient given by
\begin{equation}
  a_{j,k} = \langle V, \psi_{j,k} \rangle = \int_{-\infty}^{\infty}
  V(x') \psi^{*}_{j,k}(x') dx'
\end{equation}
quantifies the magnitude of fluctuation of $V(x)$ around this
space-scale point.
Hereafter we use Meyer's wavelet. It is a real analytic function that
decays faster than any power function as $|x| \to \infty$, and its
moment of any order vanishes, i.e.,
\begin{equation}
  \int_{-\infty}^{\infty} x^n \psi(x) dx = 0, \;\;\; (n \geq 0).
\end{equation}
Further, it has a smooth Fourier transform with a compact support.
These features are very preferable from the viewpoint of physicists.
For the details of Meyer's wavelet and efficient numerical algorism,
see Refs.~[Yamada \& Ohkitani, 1991; Mallat, 1999].
%%~\cite{Mallat,Yamada}.
%%
Meyer's mother wavelet and one of its child wavelet are shown in
Fig.~\ref{Fig:07}.

In the practical numerical calculation, the amplitude pattern is a
periodic function on $[0,1)$ represented by $N=2^n$ discrete
points. Therefore, we extend the pattern over the whole real axis by a
periodic extrapolation in order to apply the wavelet
transform. Consequently, in the actual numerical calculation, the
first subscript $j$ in the summation of
Eq.~(\ref{Eq:WaveletDecomposition}) indicating the scale runs from $0$
to $n-1$, and the second subscript $k$ indicating the translation runs
from $0$ to $2^j-1$.

\begin{figure}[htbp]
  \begin{center}
    \leavevmode
    \epsfxsize=0.4\textwidth
    \epsfbox{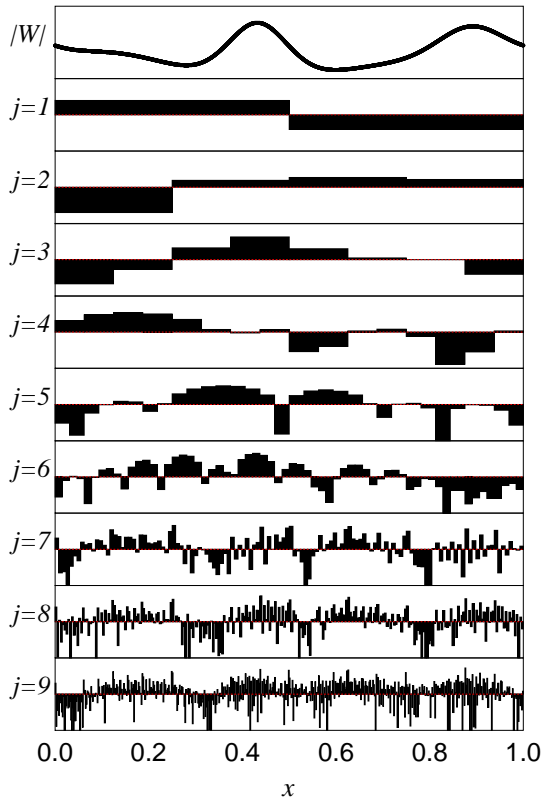}
    \caption{Coarse-grained field $b_j(x)$ obtained from the logarithm
      of the squared wavelet coefficient of the LCGL equation. The top
      panel shows the original amplitude pattern $|W(x)|$. The scale
      $j$ ranges from $1$ to $9$ from above. Each graph is normalized
      with its mean and variance. $N=2^{10}$ modes are used in the
      numerical calculation.}
    \label{Fig:08}
  \end{center}
\end{figure}

\begin{figure}[htbp]
  \begin{center}
    \leavevmode
    \epsfxsize=0.4\textwidth
    \epsfbox{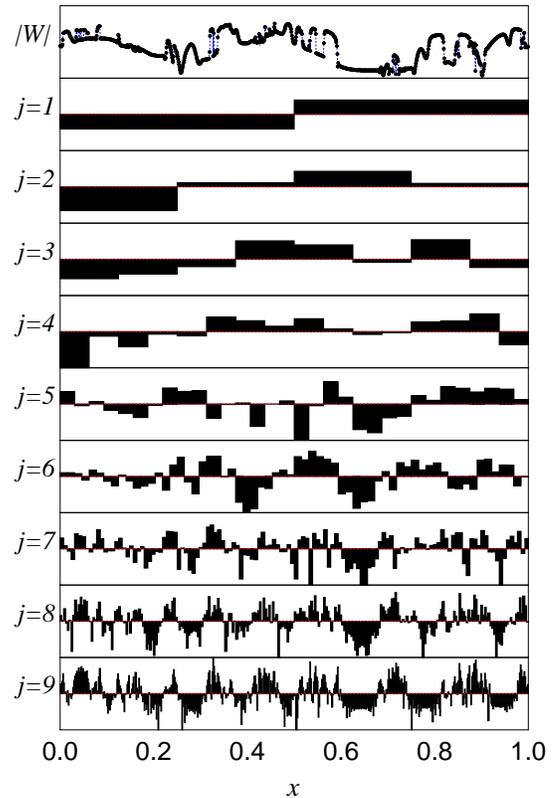}
    \caption{Coarse-grained field $b_j(x)$ obtained from the logarithm
      of the squared wavelet coefficient of the NCGL equation. The top
      panel shows the original amplitude pattern $|W(x)|$. The scale
      $j$ ranges from $1$ to $9$ from above. Each graph is normalized
      with its mean and variance. $N=2^{10}$ oscillators are used in
      the numerical calculation.}
    \label{Fig:09}
  \end{center}
\end{figure}

Rather than using the obtained expansion coefficient directly, we
define a new coarse-grained field $b_{j}(x) \; (0 \leq x < 1)$ by a
piecewise constant interpolation from the expansion coefficient
$a_{j,k}$ as
\begin{equation}
  b_{j}(x) = \log |a_{j,k}|^2 \;\; \left( \frac{k}{2^j} \leq x <
    \frac{k+1}{2^j}, \;\; k=0, 1, ..., 2^{j}-1 \right).
\end{equation}
Here we take a square of the coefficient, because we are interested in
the absolute intensity of the fluctuation at a certain position and
scale.
Further taking the logarithm is merely for convenience sake here; we
obtain similar results without taking the logarithm. However, there
also exists some discussion claiming that taking the logarithm is more
appropriate when the system under consideration possesses a random
multiplicative process [Arn\'eodo {\it et al.}, 1998a; Arn\'eodo {\it
  et al.}, 1998b],
%%\cite{Arneodo1,Arneodo2},
and the NCGL equation is actually considered this case.
Figures~\ref{Fig:08} and \ref{Fig:09} display the coarse-grained
fields $b_{j}(x)$ for the LCGL and NCGL equations. The obtained data
for the LCGL equation are very noisy for large $j$ values, since
components whose wavelengths are shorter than the dissipation scale
$\sqrt{D} \simeq 1/16$ decay very quickly. On the other hand, we can
observe clear positional correlation between scales down to a very
small wavelength in the case of the NCGL equation.

Let $V_1(x)$ be the amplitude pattern at time $t$, i.e., $V_1(x) =
|W(x,t)|$, and $V_2(x)$ be the amplitude pattern evolved from $V_1(x)$
for a period of $\Delta t$, i.e., $V_2(x) = |W(x,t + \Delta t)|$.
From these amplitude patterns, we obtain coarse-grained fields
$b^{1}_{j_1}(x)$ and $b^{2}_{j_2}(x)$ at scales $j_1$ and $j_2$.
We then define deviations of these fields from their mean values as
\begin{equation}
  \Delta b^{1}_{j_1}(x) = b^{1}_{j_1}(x) - \overline{ ( b^{1}_{j_1}, I
    ) },
\end{equation}
and
\begin{equation}
  \Delta b^{2}_{j_2}(x) = b^{2}_{j_2}(x) - \overline{ ( b^{2}_{j_2}, I
    ) },
\end{equation}
where $I$ is a constant-valued function $I(x) \equiv 1$, and the inner product 
of $f(x)$ and $g(x)$ is defined as
\begin{equation}
  ( f, g ) = \int_{0}^{1} f(x) g(x) dx.
  \label{Eq:InnerProduct}
\end{equation}
Now we define a kind of cross-correlation matrix $C_{j_1, j_2}(\Delta
t)$ from these $\Delta b^{1}_{j_1}(x)$ and $\Delta b^{2}_{j_2}(x)$ as
\begin{equation}
  C_{j_1, j_2}(\Delta t) = \frac{
    \overline{ ( \Delta b^{1}_{j_1},\; \Delta b^{2}_{j_2} ) }
    }{
    \left[\;
      \overline{ ( \Delta b^{1}_{j_1},\; \Delta b^{1}_{j_1} ) }
      \;
      \overline{ ( \Delta b^{2}_{j_2},\; \Delta b^{2}_{j_2} ) }
    \;\right]^{1/2}
  }.
\end{equation}
From this definition, $C_{j_1, j_2}(\Delta t) = C_{j_2, j_1}(- \Delta
t)$ obviously follows.
The overlines in the above equations indicate temporal average,
assuming the stationarity of the spatio-temporal chaos under
consideration.

When $\Delta t = 0$, $V_2(x)$ represents the same amplitude pattern as
$V_1(x)$. Thus $\Delta b^{1}_{j_1}(x)$ and $\Delta b^{2}_{j_2}(x)$ are
identical, and $C_{j_1, j_2}$ is merely a symmetric matrix.
When $\Delta t \neq 0$, $V_2(x)$ differs from $V_1(x)$ to some degree
depending on $\Delta t$. Therefore $C_{j_1, j_2}$ is no longer
symmetric, and its asymmetry is expected to characterize the variation
of spatio-temporal correlation between scales.
In order to see this asymmetry clearly, let us decompose the matrix
$C_{j_1, j_2}$ into symmetric and antisymmetric parts as
\begin{equation}
  D_{j_1, j_2}(\Delta t) = \frac{1}{2} \left\{ C_{j_1, j_2}(\Delta t)
    + C_{j_2, j_1}(\Delta t) \right\},
\end{equation}
and
\begin{equation}
  E_{j_1, j_2}(\Delta t) = C_{j_1, j_2}(\Delta t) - C_{j_2,
    j_1}(\Delta t).
\end{equation}
The symmetric part $D_{j_1, j_2}$ is invariant under time-reversal
transform $\Delta t \leftrightarrow -\Delta t$, while the
antisymmetric part $E_{j_1, j_2}$ is not invariant
(antisymmetric). Thus the matrix $E_{j_1, j_2}$ is expected to
quantify the variation of spatio-temporal correlation between scales
that is not symmetric to time-reversal, namely, to give a certain
measure of the flow of information or correlation between scales.

\begin{figure}[htbp]
  \begin{center}
    \leavevmode
    \epsfxsize=0.4\textwidth
    \epsfbox{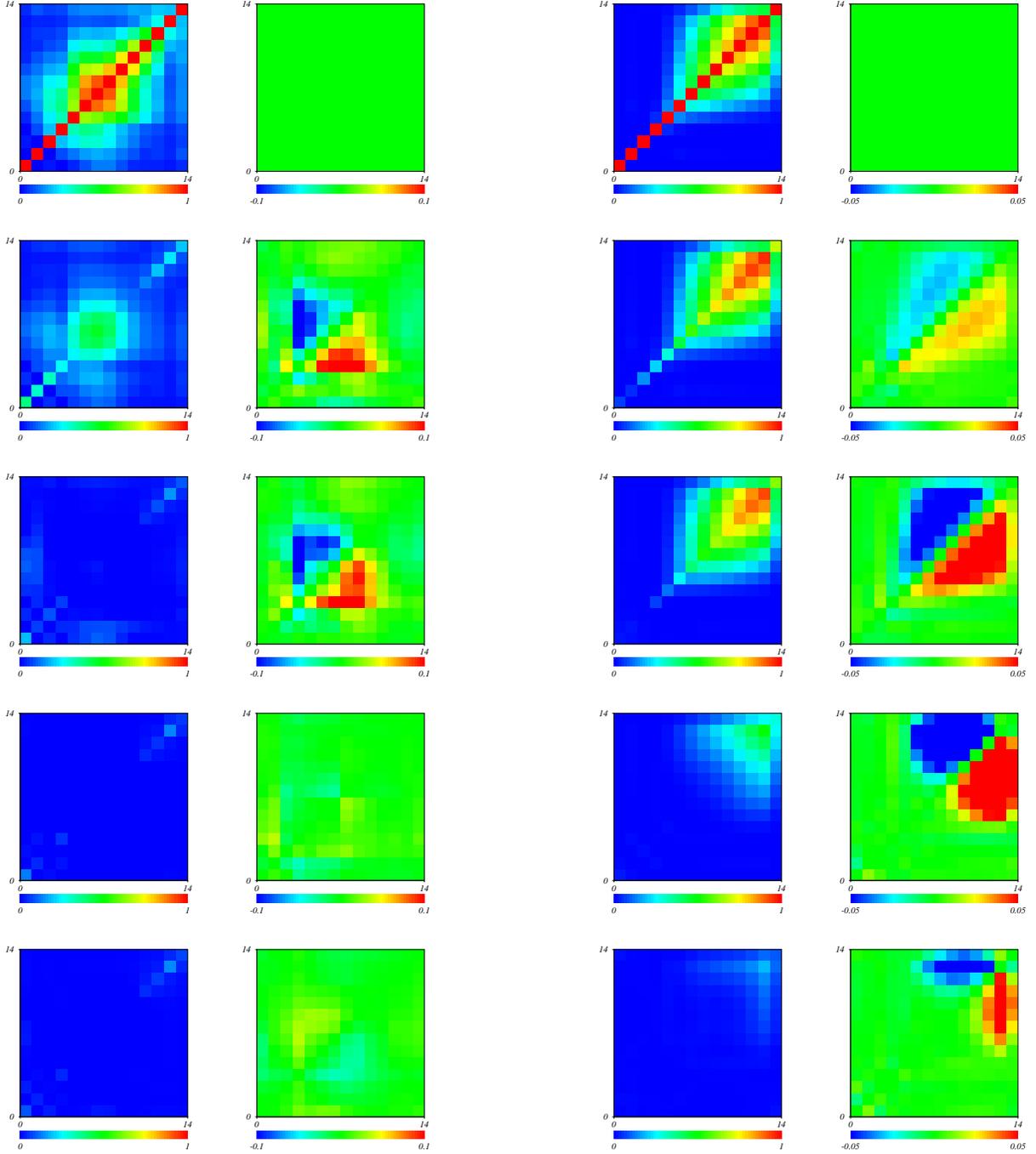}
    \caption{Evolution of the symmetric part $D_{j_1, j_2}$ (left column)
      and the antisymmetric part $E_{j_1, j_2}$ (right column) of the
      correlation matrix with the time difference $\Delta t$ for the
      LCGL equation. In each figure, the vertical axis indicates the
      scale $j_1$ at the reference time, and the horizontal axis
      indicates the scale $j_2$ at the time $\Delta t$ after the
      reference time. The scales of both axes range from $0$ to $13$.
      Time intervals between patterns are $\Delta t = 0, 1, 2, 4, 8$
      from the top. $N=2^{14}$ modes are used in the numerical
      calculation.
      The color scale of $D_{j_1, j_2}$ is from $0$ (blue) to $1$
      (red), while that of $E_{j_1, j_2}$ is from $-0.1$ (blue) to
      $0.1$ (red).}
    \label{Fig:10}
  \end{center}
\end{figure}

In Fig.~\ref{Fig:10}, the symmetric part $D_{j_1, j_2}$ and the
antisymmetric part $E_{j_1, j_2}$ of the correlation matrix obtained
for the LCGL equation are displayed using color codes for several
values of the time difference $\Delta t$.
The top row displays results obtained for the same time ($\Delta t =
0$), and the lower rows display results for larger values of the time
difference $\Delta t$.  The vertical axis of each graph represents the
scale $j_1$ of the pattern at the reference time, and the horizontal
axis represents the scale $j_2$ of the pattern $\Delta t$ after the
reference time.
First, for $\Delta t=0$, the antisymmetric part is merely zero and
only the symmetric part takes positive values. The diagonal components
naturally take largest values, since the self-correlation of the
fluctuation at the same scale is strongest. Besides, it can be seen
that there exists rather strong correlation between adjacent scales in
the dissipation region determined by the diffusion constant $D$ (the
region with the scale slightly smaller than $j \simeq 4$).
When the time difference $\Delta t$ becomes a little larger, the
symmetric part diminishes. But still the diagonal components and the
dissipation region maintain higher correlations than the others. Now,
if we look at the antisymmetric part, there appear small localized
regions in the neighborhood of the dissipation region across the
diagonal, which have negative and positive correlations, respectively.
This implies that fluctuation at a certain scale at the reference time
(vertical axis) possesses relatively strong correlation to fluctuation
at slightly smaller scale $\Delta t$ after the reference time
(horizontal axis). In this case, it can be considered as visualizing
the dissipation due to the diffusion term.
As the time difference $\Delta t$ becomes further large, both the
symmetric and antisymmetric parts tend to take smaller values, and
correlation between the patterns vanishes.
The intensity of the antisymmetric part is typically $20\%$ of the
symmetric part at its maximum.

\begin{figure}[htbp]
  \begin{center}
    \leavevmode
    \epsfxsize=0.4\textwidth
    \epsfbox{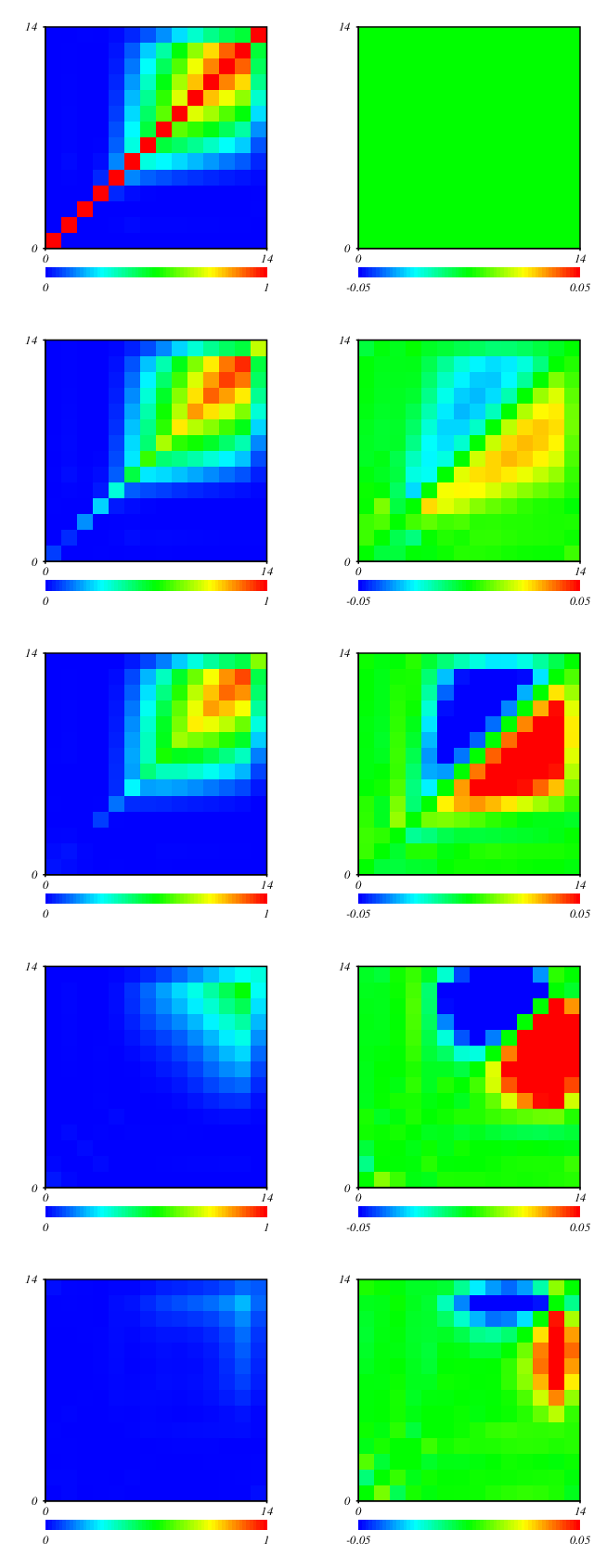}
    \caption{Evolution of the symmetric part $D_{j_1, j_2}$ (left column)
      and the antisymmetric part $E_{j_1, j_2}$ (right column) of the
      correlation matrix with the time difference $\Delta t$ for the
      NCGL equation. In each figure, the vertical axis indicates the
      scale $j_1$ at the reference time, and the horizontal axis
      indicates the scale $j_2$ at the time $\Delta t$ after the
      reference time.  The scales of both axes range from $0$ to $13$.
      Time intervals between patterns are $\Delta t = 0, 2, 5, 20, 40$
      from the top. $N=2^{14}$ oscillators are used in the numerical
      calculation.
      The color scale of $D_{j_1, j_2}$ is from $0$ (blue) to $1$
      (red), while that of $E_{j_1, j_2}$ is from $-0.05$ (blue) to
      $0.05$ (red).}
    \label{Fig:11}
  \end{center}
\end{figure}

Figure~\ref{Fig:11} displays the symmetric part $D_{j_1, j_2}$ and the
antisymmetric part $E_{j_1, j_2}$ of the correlation matrix obtained
for the NCGL equation for several values of the time difference
$\Delta t$ using color codes, as in the case of Fig.~\ref{Fig:10}.
When there is no time difference ($\Delta t=0$), only the symmetric
part takes finite values, and its diagonal components take largest
values as in the case of the LCGL equation. However, the region where
the correlation between adjacent scales takes its peak value is
located at much smaller scale.
When the time difference $\Delta t$ becomes a little larger, the
difference becomes more remarkable; there appear wide coherent regions
with positive and negative correlation, which spread over scales
shorter than the characteristic length of the system ($j \simeq 4$)
determined by the coupling range $\gamma$.
This asymmetry is maintained up to a considerably large value of
$\Delta t$, which indicates that long-wavelength components at the
reference time maintain strong correlations widely to
shorter-wavelength components at later times.
As we increase the time difference $\Delta t$ further, the correlation
between scales gradually vanishes from the long-wavelength region. But
it takes much longer than the case of the LCGL equation.
These facts suggest the existence of cascade-like propagation of some
quantity from long wavelength to shorter wavelength.
The maximum intensity of the antisymmetric part is about $10\%$ of the
symmetric part.


\newpage

\begin{figure}[htbp]
  \begin{center}
    \leavevmode
    \epsfxsize=0.4\textwidth
    \epsfbox{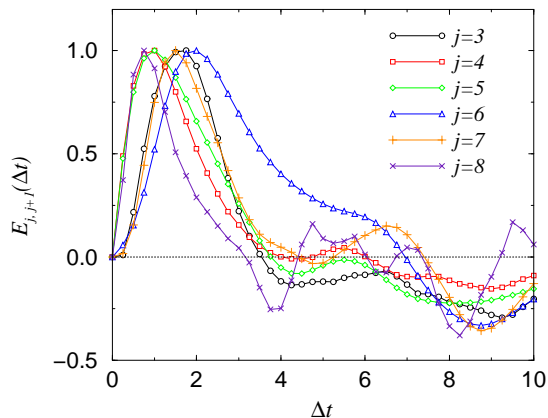}
    \caption{Antisymmetric part of the correlation coefficient $E_{j,j+1}$
      between two successive scales $j$ and $j+1$ as a function of the
      time interval $\Delta t$ for the LCGL equation. Each curve is
      normalized by its maximum value.}
    \label{Fig:12}
  \end{center}
\end{figure}

\begin{figure}[htbp]
  \begin{center}
    \leavevmode
    \epsfxsize=0.4\textwidth
    \epsfbox{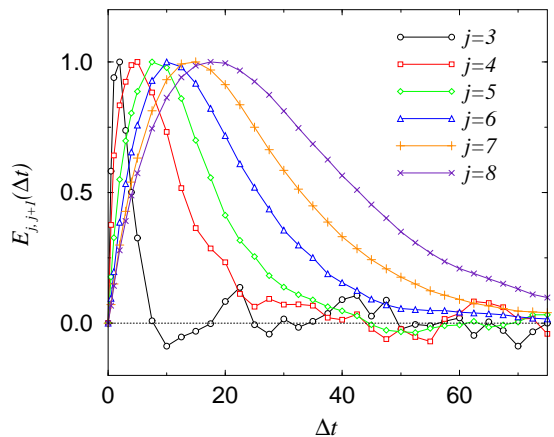}
    \caption{Antisymmetric part of the correlation coefficient $E_{j,j+1}$
      between two successive scales $j$ and $j+1$ as a function of the
      time interval $\Delta t$ for the NCGL equation. Each curve is
      normalized by its maximum value.}
    \label{Fig:13}
  \end{center}
\end{figure}

In order to see the temporal evolution of the antisymmetric part of
the correlation matrix in more detail, antisymmetric components
$E_{j,j+1}$ between adjacent scales $j$ and $j+1$ normalized by the
maximum value at each scale are displayed in Figs.~\ref{Fig:12} and
\ref{Fig:13} for scales smaller than the characteristic length.
In the case of the LCGL equation, there is no clear ordering of the
temporal variation of correlation. Each curve has its maximum at some
small value of $\Delta t$, and vanishes quickly.
In the case of the NCGL equation, on the other hand, there exists a
clear time ordering of the curves in scale; the correlation between
long-wavelength fluctuation (small $j$ values) becomes large earlier
and decreases quickly, while the correlation between short-wavelength
fluctuation (large $j$ values) becomes large later and decreases very
slowly.  (Note the difference of time scale between the graphs for the
LCGL and NCGL equations.)


\section{Discussion}

As we demonstrated, our simple method based on the correlation matrix
seems to visualize the spatio-temporal correlation between scales of
spatio-temporal chaos successfully.
Our method clearly visualized the difference of dynamical process
between two systems with different natures. Especially, for the
non-locally coupled system, it verified the existence of cascade-like
propagation of correlation from long-wavelength components to
shorter-wavelength components.

First, the use of the wavelet transform is essential for our
results. It seems difficult to obtain similar results using a method
that flattens out the spatial information completely, like the Fourier
transform.  As can clearly be seen from the definition of the inner
product given in Eq.~(\ref{Eq:InnerProduct}), our method captures the
simultaneous fluctuation of two fields at the same position in space.
For example, if we use a different definition of the inner product
that loses the spatial information by taking spatial average of each
field first, we cannot observe the correlation between scales so
clearly as in Figs.~\ref{Fig:10} and \ref{Fig:11}.
Similarly, if we change the definition of the inner product so as to
multiply the fields with their origins shifted by half the system
size, the correlation matrix almost vanishes.
(This tendency is more remarkable for the NCGL equation. In the case
of the LCGL equation, though not so clear, we obtain similar
asymmetric correlation matrices with these modified definitions of the
inner product. This is because with the values of $D$ and $L$ we used
in our calculation, the ratio of the diffusion constant to the system
size is considerably large, and some coherence over the whole system
still remains. If we further decrease the value of $D$, the
correlation matrix tends to vanish as in the case of the NCGL
equation.)

Conversely, the fact that we succeeded in visualization using the
definition of the inner product given in Eq.~(\ref{Eq:InnerProduct})
implies that the cascade process occurs in spatially localized regions
in our system.
However, it is frequently seen in other systems that the spatial
structure of the pattern collapses while moving constantly, as in the
case of convective instability. In order to detect such phenomena, it
would be possible to generalize the definition of the inner product in
such a way that the distance between the origins of the fields is
increased with $\Delta t$.

%We would like to stress that though we used Meyer's wavelet in this
%paper, it is just a matter of taste. We believe that we will obtain
%essentially the same result with other types of wavelets. We are not
%at all interested in non-generic results that strongly depend on the
%details of the wavelets used in calculation.

Precisely speaking, the result visualized by our method is not a flow
of causality, but merely a flow of correlation. Namely, it indicates
that fluctuation at the reference time at some scale varies in unison
with fluctuation at the later time at some other scale, but it does
not necessarily mean that the former one actually affects the latter
one.
Of course, it is natural to interpret it as causality in our cases.
But in order to be exact, it will be possible to perturb a certain
mode at some scale, e.g. by using a sinusoidal wave, and visualize the
propagation of its aftereffect to prove that it is indeed a causal
relationship.

Finally, though we did not give a detailed discussion in this paper,
we will be able to investigate the dynamical process of the
spatio-temporal chaos of non-locally coupled oscillators in more
detail by quantitatively analyzing the results obtained by our method.
For example, from the peak positions of the $E_{j,j+1}$ curves shown
in Fig.\ref{Fig:13}, we will be able to study the dependence of
characteristic time scale of the underlying cascade process on the
coupling strength. Such a study will also be an interesting future
subject.

\acknowledgments{H. N. gratefully acknowledges M. Hayashi for useful
  discussion, and S. Amari for providing an excellent environment for
  scientific study. He also thanks K. Ishioka, Y. Taniguchi, C. Liu,
  S. Kato, Y. Kitano, and N. Kobayashi for their warm hospitality
  during his stay in University of Tokyo, and T. Takami and
  T. Mizuguchi for providing a nice visualization tool. This work is
  partly supported by the JSPS research fellowship for young
  scientists, and partly by RIKEN.}

\end{document}